# From Theory to Practice: Sub-Nyquist Sampling of Sparse Wideband Analog Signals

Moshe Mishali, *Student Member, IEEE*, and Yonina C. Eldar, *Senior Member, IEEE*

*Abstract*—Conventional sub-Nyquist sampling methods for analog signals exploit prior information about the spectral support. In this paper, we consider the challenging problem of blind sub-Nyquist sampling of multiband signals, whose unknown frequency support occupies only a small portion of a wide spectrum. Our primary design goals are efficient hardware implementation and low computational load on the supporting digital processing. We propose a system, named the modulated wideband converter, which first multiplies the analog signal by a bank of periodic waveforms. The product is then lowpass filtered and sampled uniformly at a low rate, which is orders of magnitude smaller than Nyquist. Perfect recovery from the proposed samples is achieved under certain necessary and sufficient conditions. We also develop a digital architecture, which allows either reconstruction of the analog input, or processing of any band of interest at a low rate, that is, without interpolating to the high Nyquist rate. Numerical simulations demonstrate many engineering aspects: robustness to noise and mismodeling, potential hardware simplifications, realtime performance for signals with time-varying support and stability to quantization effects. We compare our system with two previous approaches: periodic nonuniform sampling, which is bandwidth limited by existing hardware devices, and the random demodulator, which is restricted to discrete multitone signals and has a high computational load. In the broader context of Nyquist sampling, our scheme has the potential to break through the bandwidth barrier of state-of-the-art analog conversion technologies such as interleaved converters.

## I. Introduction

**R**ADIO frequency (RF) technology enables the modulation of narrowband signals by high carrier frequencies. Consequently, manmade radio signals are often sparse. That is, they consist of a relatively small number of narrowband transmissions spread across a wide spectrum range. A convenient way to describe this class of signals is through a multiband model. The frequency support of a multiband signal resides within several continuous intervals spread over a wide spectrum. Figure 1 depicts a typical communication application, the wideband receiver, in which the received signal follows the multiband model. The basic operations in such an application are conversion of the incoming signal to digital, and low-rate processing of some or all of the individual transmissions. Ultimately, the digital product is transformed back to the analog domain for further transmission.

Due to the wide spectral range of multiband signals, their Nyquist rates may exceed the specifications of the best analog-to-digital converters (ADCs) by orders of magnitude. Any

Moshe Mishali and Yonina C. Eldar are with the Technion—Israel Institute of Technology, Haifa Israel. Emails: moshiko@tx.technion.ac.il, yonina@ee.technion.ac.il.

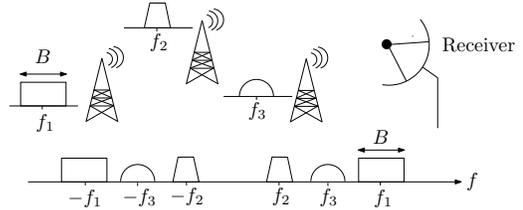

Fig. 1. Three RF transmissions with different carriers $f_i$. The receiver sees a multiband signal (bottom drawing).

attempt to acquire a multiband signal must therefore exploit its structure in an intelligent way. When the carrier frequencies are known, a common practical engineering approach is to demodulate the signal by its carrier frequency such that the spectral contents of a band of interest are centered around the origin. A lowpass filter follows in order to reject frequencies due to the other bands. Conversion to digital is then performed at a rate matching the actual information width of the band of interest. Repeating the process for each band separately results in a sampling rate which is the sum of the bandwidths. This method achieves the minimal sampling rate, as derived by Landau [1], which is equal to the actual frequency occupancy. An alternative sampling approach that does not require analog preprocessing was proposed in [2]. In this strategy, periodic nonuniform sampling is used to directly sample a multiband signal at an average rate approaching that derived by Landau. Both conventional demodulation and the method of [2] rely on knowledge of the carrier frequencies.

In scenarios in which the carrier frequencies are unknown to the receiver, or vary in time, a challenging task is to design a *spectrum-blind* receiver at a sub-Nyquist rate. In [3], [4] a multicoset sampling strategy was developed, independent of the signal support, to acquire multiband signals at low rates. Although the sampling method is blind, in order to recover the original signal from the samples, knowledge of the frequency support is needed. Recently [5], we proposed a fully spectrum-blind system based on multicoset sampling. Our system does not require knowledge of the frequency support in either the sampling or the recovery stages. To reconstruct the signal blindly, we developed digital algorithms that process the samples and identify the unknown spectral support. Once the support is found, the continuous signal is reconstructed using closed-form expressions.

Periodic nonuniform sampling is a popular approach in the broader context of analog conversion when the spectrum is fully occupied. Instead of implementing a single ADC at a high-rate $R$, interleaved ADCs use $M$ devices at rate $R/M$ with appropriate time shifts [6]–[8]. However, time interleaving has two fundamental limitations. First, the $M$ lowrate samplers have to share an analog front-end which must tolerate the input bandwidth $R$. With today's technology the possible front-ends are still far below the wideband regime. Second, maintaining accurate time shifts, on the order of $1/RM$, is difficult to implement. Multicoset sampling, is a special case of interleaved ADCs, so that the same limitations apply. In Section II-B we discuss in more detail the difficulty in implementing interleaved ADCs and multicoset sampling. In practice, such systems are limited to intermediate input frequencies and cannot deal with wideband inputs.

Recently, a new architecture to acquire multitone signals, called the random demodulator, was studied in the literature of compressed sensing (CS) [9], [10]. In this approach, the signal is modulated by a high-rate pseudorandom number generator, integrated, and sampled at a low rate. This scheme applies to signals with finite set of harmonics chosen from a fixed uniform grid. Time-domain analysis shows that CS algorithms can recover such a multitone signal from the proposed samples [10]. However, as discussed in Section VI, truly analog signals require a prohibitively large number of harmonics to approximate them well within the discrete model, which in turn renders the reconstruction computationally infeasible and very sensitive to the grid choice. Furthermore, the time-domain approach precludes processing at a low rate, even for multitone inputs, since interpolation to the Nyquist rate is an essential ingredient in the reconstruction.

In this paper, we aim to combine the advantages of the previous approaches: The ability to treat analog multiband models, a sampling stage with a practical implementation, and a spectrum-blind recovery stage which involves efficient digital processing. In addition, we would like a method that allows low-rate processing, namely the ability to process any one of the transmitted bands without first requiring interpolation to the high Nyquist rate.

Our main contribution is an analog system, referred to as the modulated wideband converter (MWC), which is comprised of a bank of modulators and lowpass filters. The signal is multiplied by a periodic waveform, whose period corresponds to the multiband model parameters. A square-wave alternating at the Nyquist rate is one choice; other periodic waveforms are also possible. The goal of the modulator is to alias the spectrum into baseband. The modulated output is then lowpass filtered, and sampled at a low rate. The rate can be as low as the expected width of an individual transmission. Based on frequency-domain arguments, we prove that an appropriate choice of the parameters (waveform period, sampling rate) guarantees that our system uniquely determines a multiband input signal. In addition, we describe how to trade the number of channels by a higher rate in each branch, at the expense of additional processing. Theoretically, this method allows to collapse the entire system to a single channel operating at a rate lower than Nyquist.

Our second contribution is a digital architecture which enables processing of the samples for various purposes. Reconstruction of the original analog input is one possible function. Perhaps more useful is the capability of the proposed system to generate lowrate sequences corresponding to each of the bands, which, in principle, allow subsequent digital processing of each band at a low rate. This architecture also has the ability to treat inputs with time-varying support. At the heart of the digital processing lies the continuous to finite (CTF) block from our previous works [5], [11]. The CTF separates the support recovery from the rest of the operations in the digital domain. In our previous works, the CTF required costly digital processing at the Nyquist rate, and therefore provided only analog reconstruction at the price of high rate computations. In contrast, here, the CTF computations are carried out directly on the lowrate samples.

The main theme of this paper is going from theory to practice, namely tying together a theoretical sampling approach with practical engineering aspects. Besides the uniqueness theorems and stability conditions, we make use of extensive numerical simulations, in Section V, to study typical wideband scenarios. The simulations demonstrate robustness to noise and signal mismodeling, potential hardware simplifications in order to reduce the number of devices, fast adaption to time-varying spectral support, and the performance with quantized samples. A circuit-level implementation of the MWC is under development and will be reported in [12].

The paper is organized as follows. Section II describes the multiband model and points out limitations of multicoset sampling in the wideband regime. In Section III, we describe the MWC system and provide a frequency-domain analysis of the resulting samples. This leads to a concrete parameter selection which guarantees a unique signal matching the digital samples. We conclude the section with a discussion on the tradeoff between the number of channels, rate, and complexity. The architecture for low-rate processing and recovery, is presented in Section IV. In Section V, we conduct a detailed numerical evaluation of the proposed system. A review of related work concludes the paper in Section VI.

## II. FORMULATION AND BACKGROUND

### A. Problem formulation

Let $x(t)$ be a real-valued continuous-time signal in $L_2$. Throughout the paper, continuous signals are assumed to be bandlimited to $\mathcal{F} = [-1/2T, +1/2T]$. Formally, the Fourier transform of $x(t)$, which is defined by

$$X(f) = \int_{-\infty}^{\infty} x(t)e^{-j2\pi ft} \mathrm{d}t, \tag{1}$$

is zero for every $f \notin \mathcal{F}$. We denote by $f_{\text{NYQ}} = 1/T$ the Nyquist rate of $x(t)$. For technical reasons, it is also assumed that $X(f)$ is piecewise continuous in $f$. We treat signals from the multiband model $\mathcal{M}$ defined below.

*Definition 1:* The set $\mathcal{M}$ contains all signals $x(t)$, such that the support of the Fourier transform $X(f)$ is contained within a union of $N$ disjoint intervals (bands) in $\mathcal{F}$, and each of the bandwidths does not exceed $B$.

Signals in $\mathcal{M}$ have an even number $N$ of bands due to the conjugate symmetry of $X(f)$. The band positions are arbitrary, and in particular, unknown in advance. A typical spectral support of a signal from the multiband model $\mathcal{M}$ is illustrated in the example of Fig. 1, in which $N = 6$ and $B, f_{\text{NYQ}}$ are dictated by the specifications of the possible transmitters.

We wish to design a sampling system for signals from the model $\mathcal{M}$ that satisfies the following properties:

1) The sampling rate should be as low as possible;
2) the system has no prior knowledge of the band locations;
3) the system can be implemented with existing analog devices and (preferably low-rate) ADCs.

Together with the sampling stage we need to design a reconstruction scheme, which converts the discrete samples back to the continuous-time domain. This stage may involve digital processing prior to reconstruction. An implicit (but crucial) requirement is that recovery involves a reasonable amount of computations. Realtime applications may also necessitate short latency from input to output and a constant throughput. Therefore, two main factors dictate the input spectrum range that the overall system can handle: analog hardware at the required rate that can convert the signal to digital, and a digital stage that can accommodate the computational load.

In our previous work [5], we proved that the minimal sampling rate for $\mathcal{M}$ to allow perfect blind reconstruction is $2NB$, provided that $2NB$ is lower than the Nyquist rate. The case $2NB \geq f_{\text{NYQ}}$ represents signals which occupy more than half of the Nyquist range. No rate improvement is possible in that case (for arbitrary signals), and thus we assume $2NB < f_{\text{NYQ}}$ in the sequel. Concrete algorithms for blind recovery, achieving the minimal rate, were developed in [5] based on a multicoset sampling strategy. The next section briefly describes this method, which achieves the goals of minimal rate and blindness. However, limitations of practical ADCs, which we detail in the next section, render multicoset sampling impractical for wideband signals. As described later in Section III-A, the sampling scheme proposed in this paper circumvents these limitations and has other advantages in terms of practical implementation.

### B. Multicoset using practical ADCs

In multicoset sampling, samples of $x(t)$ are obtained on a periodic and nonuniform grid which is a subset of the Nyquist grid. Formally, denote by $x(nT)$ the sequence of samples taken at the Nyquist rate. Let $M$ be a positive integer, and $C = \{c_i\}_{i=1}^m$ be a set of $m$ distinct integers with $0 \leq c_i \leq M-1$. Multicoset samples consist of $m$ uniform sequences, called cosets, with the $i$th coset defined by

$$x_{c_i}[n] = x(nMT + c_iT), \quad n \in \mathbb{Z}. \tag{2}$$

Only $m < M$ cosets are used, so that the average sampling rate is $m/MT$, which is lower than the Nyquist rate $1/T$.

A possible implementation of the sampling sequences (2) is depicted in Fig. 2-a. The building blocks are $m$ uniform samplers at rate $1/MT$, where the $i$th sampler is shifted by $c_iT$ from the origin. Although this scheme seems intuitive and straightforward, practical ADCs introduce an inherent bandwidth limitation, which distorts the samples. The distortion mechanism, which is modeled as a preceding lowpass filter in Fig. 2-b, becomes crucial for high rate inputs. To understand this phenomenon, we focus on the model of a practical ADC, Fig. 2-b, ignoring the time shifts for the moment. A uniform ADC at rate $r$ samples/sec attempts to output pointwise samples of the input. The design process and manufacturing technology result in an additional property, termed analog (full-power) bandwidth [13], which determines the maximal frequency $b$ that the device can handle. Any spectral content beyond $b$ Hz is attenuated and distorted. The bandwidth limitation $b$ is inherent and cannot be separated from the ADC. Therefore, manufacturers usually recommend adding a preceding external anti-aliasing lowpass filter, with cutoff $b$, since the internal one has a parasitic response. The ratio $b/r$ affects the complexity of the ADC circuit design, and is typically in the range [14]

$$1.5r \leq b \leq 7r. \tag{3}$$

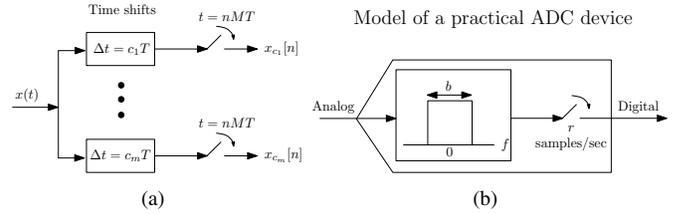

Fig. 2. Schematic implementation of multicoset sampling (a) requires no filtering between the time shifts and the actual sampling. However, the front-end of a practical ADC has an inherent bandwidth limitation, which is modeled in (b) as a lowpass filter preceding the uniform sampling.

The practical ADC model raises two difficulties in implementing multicoset sampling. First, RF technology allows transmissions at rates which exceed the analog bandwidth $b$ of state-of-the-art devices, typically by orders of magnitude. For example, ADC devices manufactured by Analog Devices Corp. have front-end bandwidths which reach up to $b =$780 MHz [14]. Therefore, any attempt to acquire a wideband signal with a practical ADC results in a loss of the spectral contents beyond $b$ Hz. The sample sequences (2) are attenuated and distorted and are no longer pointwise values of $x(t)$. This limitation is fundamental and holds in other architectures of multicoset (*e.g.*, a single ADC triggered by a nonuniform clock). The second issue is a waste of resources, which is less severe, but applies also when the Nyquist rate $f_{\text{NYQ}} = b$ for some available device. For a signal with a sparse spectrum, multicoset reduces the average sampling rate by using only $m$ out of $M$ possible cosets, where $M \gg 1$ is commonly used. Each coset in Fig. 2 samples at rate $f_{\text{NYQ}}/M$. Therefore, the ADC samples at rate $r = b/M$, which is far below the standard range (3). This implies sampling at a rate which is much lower than the maximal capability of the ADC.

As a consequence, implementing multicoset for wideband signals requires the design of a specialized fine-tuned ADC circuit, in order to meet the wide analog bandwidth, and still exploit the nonstandard ratio $b/r$ that is expected. Though this may be an interesting task for experts, it contradicts the basic goal of our design – that is, using standard and available devices. In [15] a nonconventional ADC is designed by means of high-rate optical devices. The hybrid optic-electronic system introduces a front-end whose bandwidth reaches the wideband regime, at the expense and size of an optical system. Unfortunately, at present, this performance cannot be achieved with pure electronic technology.

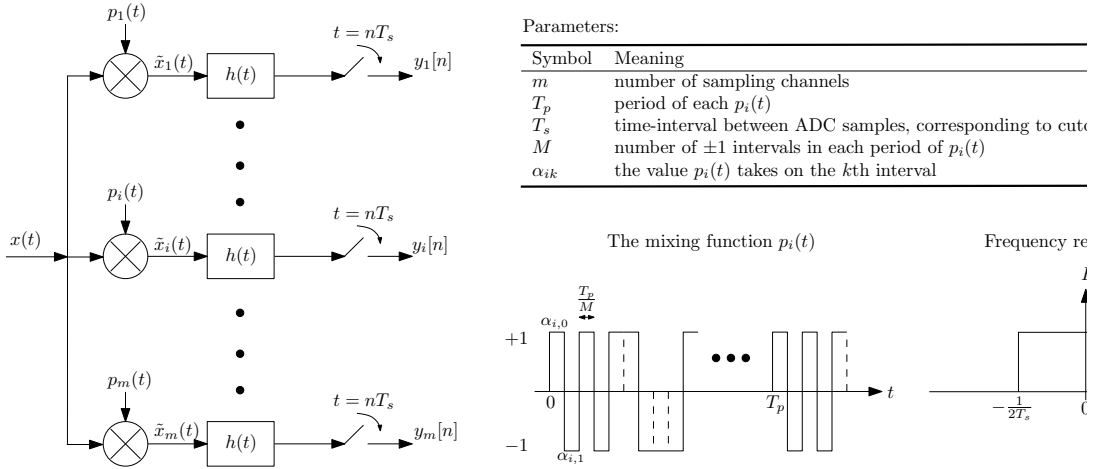

Fig. 3. The modulated wideband converter – a practical sampling stage for multiband signals.

Another practical issue of multicoset sampling, which also exists in the optical implementation, arises from the time shift elements. Maintaining accurate time delays between the ADCs in the order of the Nyquist interval $T$ is difficult. Any uncertainty in these delays influences the recovery from the sampled sequences [16]. A variety of different algorithms have been proposed in the literature in order to compensate for timing mismatches. However, this adds substantial complexity to the receiver [17], [18].

## III. SAMPLING

We now present an alternative sampling scheme that uses available devices, does not suffer from analog bandwidth issues and does not require non-zero time synchronization. The system, referred to as the modulated wideband converter (MWC), is schematically drawn in Fig. 3 with its various parameters. In the next subsections, the MWC is described and analyzed for arbitrary sets of parameters. In Section III-C, we specify a parameter choice, independent of the band locations, that approaches the minimal rate. The resulting system, which is comprised of the MWC of Fig. 3 and the recovery architecture that is presented in the next section, satisfies all the requirements of our problem formulation.

### A. System description

Our system exploits spread-spectrum techniques from communication theory [19], [20]. An analog mixing front-end aliases the spectrum, such that a spectrum portion from each band appears in baseband. The system consists of several channels, implementing different mixtures, so that, in principle, a sufficiently large number of mixtures allows to recover a relatively sparse multiband signal.

More specifically, the signal $x(t)$ enters $m$ channels simultaneously. In the $i$th channel, $x(t)$ is multiplied by a mixing function $p_i(t)$, which is $T_p$-periodic. After mixing, the signal spectrum is truncated by a lowpass filter with cutoff $1/(2T_s)$ and the filtered signal is sampled at rate $1/T_s$. The sampling rate of each channel is sufficiently low, so that existing commercial ADCs can be used for that task. The design parameters are therefore the number of channels $m$, the period $T_p$, the sampling rate $1/T_s$, and the mixing functions $p_i(t)$ for $1 \le i \le m$.

For the sake of concreteness, in the sequel, $p_i(t)$ is chosen as a piecewise constant function that alternates between the levels $\pm 1$ for each of $M$ equal time intervals. Formally,

$$p_i(t) = \alpha_{ik}, \quad k\frac{T_p}{M} \le t \le (k+1)\frac{T_p}{M}, \quad 0 \le k \le M-1, \quad (4)$$

with $\alpha_{ik} \in \{+1, -1\}$, and $p_i(t + nT_p) = p_i(t)$ for every $n \in \mathbb{Z}$. Other choices for $p_i(t)$ are possible, since in principle we only require that $p_i(t)$ is periodic.

The system proposed in Fig. 3 has several advantages for practical implementation:

(**A1**) Analog mixers are a provable technology in the wideband regime [21], [22]. In fact, since transmitters use mixers to modulate the information by a high-carrier frequency, the mixer bandwidth defines the input bandwidth.

(**A2**) Sign alternating functions can be implemented by a standard (high rate) shift register. Today's technology allows to reach alternation rates of 23 GHz [23] and even 80 GHz [24].

(**A3**) Analog filters are accurate and typically do not require more than a few passive elements (*e.g.,* capacitors and coils) [25].

(**A4**) The sampling rate $1/T_s$ matches the cutoff of $H(f)$. Therefore, an ADC with a conversion rate $r = 1/T_s$, and any bandwidth $b \ge 0.5r$ can be used to implement this block, where $H(f)$ serves as a preceding anti-aliasing filter. In the sequel, we choose $1/T_s$ on the order of $B$, which is the width of a single band of $x(t) \in \mathcal{M}$. In practice, this sampling rate allows flexible choice of an ADC from a variety of commercial devices in the low rate regime.

(**A5**) Sampling is synchronized in all channels, that is there are no time shifts. This is beneficial since the trigger for all ADCs can be generated accurately (*e.g.,* with a zero-delay synchronization device [26]). The same clock can be used for a subsequent digital processor which receives the sample sets at rate $1/T_s$.

Note that the front-end preprocessing must be carried out by analog means, since both the mixer and the analog filter operate on wideband signals, at rates which are far beyond digital processing capabilities. In fact, the mixer output $\tilde{x}_i(t)$ is not bandlimited, and therefore there is no way to replace the analog filter by a digital unit even if the converter is used for low-rate signals. The purely analog front-end is the key to overcome the bandwidth limitation of ADCs.

### B. Frequency domain analysis

We now derive the relation between the sample sequences $y_i[n]$ and the unknown signal $x(t)$. This analysis is used for several purposes in the following sections. First, for specifying a choice of parameters ensuring a unique mapping between $x(t)$ and the sequences $y_i[n]$. Second, we use this analysis to explain the reconstruction scheme. Finally, stability and implementation issues will also be based on this development. To this end, we introduce the definitions

$$f_p = 1/T_p, \quad \mathcal{F}_p = [-f_p/2, +f_p/2] \quad (5a)$$
$$f_s = 1/T_s, \quad \mathcal{F}_s = [-f_s/2, +f_s/2], . \quad (5b)$$

Consider the $i$th channel. Since $p_i(t)$ is $T_p$-periodic, it has a Fourier expansion

$$p_i(t) = \sum_{l=-\infty}^{\infty} c_{il} e^{j \frac{2\pi}{T_p} l t}, \quad (6)$$

where

$$c_{il} = \frac{1}{T_p} \int_0^{T_p} p_i(t) e^{-j \frac{2\pi}{T_p} l t} dt. \quad (7)$$

The Fourier transform of the analog multiplication $\tilde{x}_i(t) = x(t) p_i(t)$ is evaluated as

$$\begin{aligned}
\tilde{X}_i(f) &= \int_{-\infty}^{\infty} \tilde{x}_i(t) e^{-j 2 \pi f t} dt \\
&= \int_{-\infty}^{\infty} x(t) \left( \sum_{l=-\infty}^{\infty} c_{il} e^{j \frac{2\pi}{T_p} l t} \right) e^{-j 2 \pi f t} dt \\
&= \sum_{l=-\infty}^{\infty} c_{il} \int_{-\infty}^{\infty} x(t) e^{-j 2 \pi \left( f - \frac{l}{T_p} \right) t} dt \\
&= \sum_{l=-\infty}^{\infty} c_{il} X(f - l f_p). \quad (8)
\end{aligned}$$

Therefore, the input to $H(f)$ is a linear combination of $f_p$-shifted copies of $X(f)$. Since $X(f) = 0$ for $f \notin \mathcal{F}$, the sum in (8) contains (at most) $\lceil f_{\text{NYQ}}/f_p \rceil$ nonzero terms[1].

The filter $H(f)$ has a frequency response which is an ideal rectangular function, as depicted in Fig. 3. Consequently, only frequencies in the interval $\mathcal{F}_s$ are contained in the uniform sequence $y_i[n]$. Thus, the discrete-time Fourier transform

---

[1]The ceiling operator $\lceil a \rceil$ returns the greater (or equal) integer which is closest to $a$.

(DTFT) of the $i$th sequence $y_i[n]$ is expressed as

$$\begin{aligned}
Y_i(e^{j 2 \pi f T_s}) &= \sum_{n=-\infty}^{\infty} y_i[n] e^{-j 2 \pi f n T_s} \\
&= \sum_{l=-L_0}^{+L_0} c_{il} X(f - l f_p), \quad f \in \mathcal{F}_s, \quad (9)
\end{aligned}$$

where $\mathcal{F}_s$ is defined in (5b), and $L_0$ is chosen as the smallest integer such that the sum contains all nonzero contributions of $X(f)$ over $\mathcal{F}_s$. The exact value of $L_0$ is calculated by

$$-\frac{f_s}{2} + (L_0 + 1) f_p \geq \frac{f_{\text{NYQ}}}{2} \to L_0 = \left\lceil \frac{f_{\text{NYQ}} + f_s}{2 f_p} \right\rceil - 1. \quad (10)$$

Note that the mixer output $\tilde{x}_i(t)$ is not bandlimited, and, theoretically, depending on the coefficients $c_{il}$, the Fourier transform (8) may not be well defined. This technicality, however, is resolved in (9) since the filter output involves only a finite number of aliases of $x(t)$.

Relation (9) ties the known DTFTs of $y_i[n]$ to the unknown $X(f)$. This equation is the key to recovery of $x(t)$. For our purposes, it is convenient to write (9) in matrix form as

$$\mathbf{y}(f) = \mathbf{A} \mathbf{z}(f), \quad f \in \mathcal{F}_s, \quad (11)$$

where $\mathbf{y}(f)$ is a vector of length $m$ with $i$th element $y_i(f) = Y_i(e^{j 2 \pi f T_s})$. The unknown vector $\mathbf{z}(f) = [z_1(f), \cdots, z_L(f)]^T$ is of length

$$L = 2 L_0 + 1 \quad (12)$$

with

$$z_i(f) = X(f + (i - L_0 - 1) f_p), \quad 1 \leq i \leq L, f \in \mathcal{F}_s. \quad (13)$$

The $m \times L$ matrix $\mathbf{A}$ contains the coefficients $c_{il}$

$$\mathbf{A}_{il} = c_{i,-l} = c_{il}^*, \quad (14)$$

where the reverse order is due to the enumeration of $z_i(f)$ in (13). Fig. 4 depicts the vector $\mathbf{z}(f)$ and the effect of aliasing $X(f)$ in $f_p$-shifted copies for $N = 4$ bands, aliasing rate $f_p = 1/T_p \geq B$ and two sampling rates, $f_s = f_p$ and $f_s = 5 f_p$. Each entry of $\mathbf{z}(f)$ represents a frequency slice of $X(f)$ whose length is $f_s$. Thus, in order to recover $x(t)$, it is sufficient to determine $\mathbf{z}(f)$ in the interval $f \in \mathcal{F}_p$.

The analysis so far holds for every choice of $T_p$-periodic functions $p_i(t)$. Before proceeding, we discuss the role of each parameter. The period $T_p$ determines the aliasing of $X(f)$ by setting the shift intervals to $f_p = 1/T_p$. Equivalently, the aliasing rate $f_p$ controls the way the bands are arranged in the spectrum slices $\mathbf{z}(f)$, as depicted in Fig. 4. We choose $f_p \geq B$ so that each band contributes only a single nonzero element to $\mathbf{z}(f)$ (referring to a specific $f$), and consequently $\mathbf{z}(f)$ has at most $N$ nonzeros. In practice $f_p$ is chosen slightly more than $B$ to avoid edge effects. Thus, the parameter $T_p$ is used to translate the multiband prior $x(t) \in \mathcal{M}$ to a bound on the sparsity level of $\mathbf{z}(f)$. The sampling rate $f_s$ of a single channel sets the frequency range $\mathcal{F}_s$ in which (11) holds. It is clear from Fig. 4 that as long as $f_s \geq f_p$, recovering $x(t)$ from the sample sequences $y_i[n]$ amounts to recovery of $\mathbf{z}(f)$ from $\mathbf{y}(f)$, for every $f \in \mathcal{F}_p$. The number of channels $m$ determines the overall sampling rate $mf_s$ of the system. The simplest choice $f_s = f_p \simeq B$, which is presented on the left pane of Fig. 4, allows to control the sampling rate at a resolution of $f_p$. Later on, we explain how to trade the number of channels $m$ by a higher rate $f_s$ in each channel. Observe that setting $f_p, f_s$ determines $L$ by (10) and (12), which is the number of spectrum slices in $\mathbf{z}(f)$ that may contain energy for some $x(t) \in \mathcal{M}$.

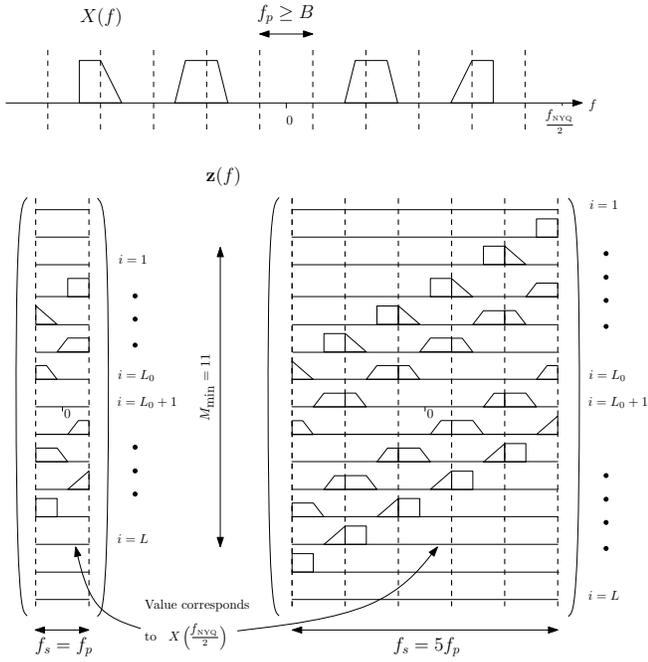

Fig. 4. The relation between the Fourier transform $X(f)$ and the vector set $\mathbf{z}(f)$ of (13). In the left pane, $f_s = f_p$ so that the length of $\mathbf{z}(f)$ is $L = 11$. The right pane demonstrates $f_s = 5f_p$ which gives $L = 15$. Entries in locations $i \leq L_0$ ($i > L_0 + 1$) contain shifted and windowed copies of $X(f)$ to the right (left) of the frequency axis. No shift occurs for the middle entry, $i = L_0 + 1$.

The role of the mixing functions appears implicitly in (11) through the coefficients $c_{il}$. Each $p_i(t)$ provides a single row in the matrix $\mathbf{A}$. Roughly speaking, $p_i(t)$ should have many transients within the time period $T_p$ so that its Fourier expansion (6) contains about $L$ dominant terms. In this case, the channel output $y_i[n]$ is a mixture of all (nonidentically zero) spectrum slices in $\mathbf{z}(f)$. The functions $p_i(t)$ should differ from each other to yield linearly independent rows in $\mathbf{A}$. The precise measure for the amount of required transients is captured by the singular values of all possible column subsets of $\mathbf{A}$ [27]. Further discussion on the choice of $p_i(t)$ appears in Section IV-D. We next study a specific choice of $p_i(t)$ – the sign waveforms.

Consider the sign alternating function $p_i(t)$, depicted in Fig. 3. Calculating the coefficients $c_{il}$ in this setting gives

$$c_{il} = \frac{1}{T_p} \int_0^{\frac{T_p}{M}} \sum_{k=0}^{M-1} \alpha_{ik} e^{-j\frac{2\pi}{T_p}l\left(t+k\frac{T_p}{M}\right)} dt$$

$$= \frac{1}{T_p} \sum_{k=0}^{M-1} \alpha_{ik} e^{-j\frac{2\pi}{M}lk} \int_0^{\frac{T_p}{M}} e^{-j\frac{2\pi}{T_p}lt} dt. \quad (15)$$

Evaluating the integral we have

$$d_l = \frac{1}{T_p} \int_0^{\frac{T_p}{M}} e^{-j\frac{2\pi}{T_p}lt} dt = \begin{cases} \frac{1}{M} & l = 0 \\ \frac{1-\theta^l}{2j\pi l} & l \neq 0 \end{cases} \quad (16)$$

where $\theta = e^{-j2\pi/M}$, and thus

$$c_{il} = d_l \sum_{k=0}^{M-1} \alpha_{ik} \theta^{lk}. \quad (17)$$

Let $\bar{\mathbf{F}}$ be the $M \times M$ discrete Fourier transform matrix (DFT) whose $i$th column is

$$\bar{\mathbf{F}}_i = \left[\theta^{0 \cdot i}, \theta^{1 \cdot i}, \cdots, \theta^{(M-1) \cdot i}\right]^T, \quad (18)$$

with $0 \leq i \leq M - 1$, and let $\mathbf{F}$ be the $M \times L$ matrix with columns $[\bar{\mathbf{F}}_{L_0} \ldots, \bar{\mathbf{F}}_{-L_0}]$ – a re-ordered column subset of $\bar{\mathbf{F}}$. Note that for $M = L$, $\mathbf{F}$ is unitary. Then, (11) can be written as

$$\mathbf{y}(f) = \mathbf{SFD}\,\mathbf{z}(f), \quad f \in \mathcal{F}_s, \quad (19)$$

where $\mathbf{S}$ is the $m \times M$ sign matrix, with $\mathbf{S}_{ik} = \alpha_{ik}$, and $\mathbf{D} = \text{diag}(d_{L_0}, \ldots, d_{-L_0})$ is an $L \times L$ diagonal matrix with $d_l$ defined by (16). As in (14), the reverse order is due to the aliasing enumeration. The dependency on the sign patterns $\{\alpha_{ik}\}$ is further expanded in (20).

A sign alternating function $p_i(t)$ is implemented by a shift register, where $M$ determines the number of flops, and $\{\alpha_{ik}\}$ initializes the shift register. The clock rate of the register $(T_p/M)^{-1}$ is also dictated by $M$. The next section shows that $M \geq L$, where $L$ is defined in (12), is one of the conditions for blind recovery. To reduce the clock rate the minimal $M$ as derived in the sequel is always preferred. Since $L$ is roughly $f_{\text{NYQ}}/B$ for $f_p = B$, this implies a large value for $M$. In practice this is not an obstacle, since standard logic gates and feedback can be used to generate a sign pattern of length $M$ (a.k.a, m-sequence) with just a few components [19], [20]. In future work, we will investigate the preferred sign pattern for stable reconstruction. In the implementation [12], we use a length $M$ register without a supporting logic, in order to allow any of the $2^M$ possible patterns.

An important consequence of periodicity is robustness to time-domain variability. As long as the waveform $p_i(t)$ is periodic, the coefficients $c_{il}$ can be computed, or can be calibrated in retrospect. Time-domain design imperfections are not important. In particular, a sign waveform whose alternations do not occur exactly on the Nyquist grid, and whose levels are not accurate $\pm 1$ levels is fine, as long as the same pattern repeats every $T_p$ seconds.

Note that the magnitude of $d_l$ decays as $l$ moves away from $l = 0$. This is a consequence of the specific choice of sign alternating waveforms for the mixing functions $p_i(t)$. Under this selection, spectrum regions of $X(f)$ are weighted according to their proximity to the origin. In the presence of noise, the signal to noise ratio depends on the band locations due to this asymmetry.

$$\underbrace{\begin{pmatrix} Y_1(e^{j2\pi f T_s}) \\ Y_2(e^{j2\pi f T_s}) \\ \vdots \\ Y_m(e^{j2\pi f T_s}) \end{pmatrix}}_{\mathbf{y}(f)} = \underbrace{\begin{bmatrix} \alpha_{1,0} & \cdots & \alpha_{1,M-1} \\ \vdots & \ddots & \vdots \\ \alpha_{m,0} & \cdots & \alpha_{m,M-1} \end{bmatrix}}_{\mathbf{S}} \underbrace{\begin{bmatrix} | & \cdots & | & \cdots & | \\ \bar{\mathbf{F}}_{L_0} & \cdots & \bar{\mathbf{F}}_0 & \cdots & \bar{\mathbf{F}}_{-L_0} \\ | & \cdots & | & \cdots & | \end{bmatrix}}_{\mathbf{F}} \underbrace{\begin{bmatrix} d_{L_0} & & \\ & \ddots & \\ & & d_{-L_0} \end{bmatrix}}_{\mathbf{D}} \underbrace{\begin{pmatrix} X(f - L_0 f_p) \\ \vdots \\ X(f) \\ \vdots \\ X(f + L_0 f_p) \end{pmatrix}}_{\mathbf{z}(f)} \quad (20)$$

---

### C. Choice of parameters

An essential property of a sampling system is that the sample sequences match a unique analog input $x(t)$, since otherwise recovery is impossible. The following theorems address this issue. The first theorem states necessary conditions on the system parameters to allow a unique mapping. A concrete parameter selection which is sufficient for uniqueness, is provided in the second theorem. The same selection works with half as many sampling channels, when the band locations are known. Thus, the system appearing in Fig. 3 can also replace conventional demodulation in the non-blind scenario. This may be beneficial for a receiver that switches between blind and non-blind modes according to availability of the transmitter carriers. More importantly, Fig. 3 suggests a possible architecture in the broader context of ADC design. The analog bandwidth of the frontend, which is dictated by the mixers, breaks the conventional bandwidth limitation in interleaved ADCs.

For brevity, we use sparsity notations in the statements below. A vector $\mathbf{u}$ is called $K$-sparse if $\mathbf{u}$ contains no more than $K$ nonzero entries. The set $\text{supp}(\mathbf{u})$ denotes the indices of the nonzeros in $\mathbf{u}$. The support of a collection of vectors over a continuous interval, such as $\mathbf{z}(\mathcal{F}_p) = \{\mathbf{z}(f) : f \in \mathcal{F}_p\}$, is defined by

$$\text{supp}(\mathbf{z}(\mathcal{F}_p)) = \bigcup_{f \in \mathcal{F}_p} \text{supp}(\mathbf{z}(f)). \quad (21)$$

A vector collection is called jointly $K$-sparse if its support contains no more than $K$ indices.

*Theorem 1 (Necessary conditions):* Let $x(t)$ be an arbitrary signal within the multiband model $\mathcal{M}$, which is sampled according to Fig. 3 with $f_p = B$. Necessary conditions to allow exact spectrum-blind recovery (of an arbitrary $x(t) \in \mathcal{M}$) are $f_s \geq f_p$, $m \geq 2N$. For mixing with sign waveforms an additional necessary requirement is

$$M \geq M_{\min} \triangleq 2 \left\lceil \frac{f_{\text{NYQ}}}{2f_p} + \frac{1}{2} \right\rceil - 1. \quad (22)$$

Note that for $f_s = f_p$, $M_{\min} = L$ of (12); see also Fig. 4.

*Proof:* Observe that according to (9) and Fig. 4, the frequency transform of the $i$th entry of $\mathbf{z}(f)$ sums $f_p$-shifted copies of $X(f)$. If $f_s < f_p$, then the sum lacks contributions from $X(f)$ for some $f \in \mathcal{F}$. An arbitrary multiband signal may contain an information band within those frequencies. Thus, $f_s \geq f_p$ is necessary.

The other conditions are necessary to allow enough linearly independent equations in (11) for arbitrary $x(t) \in \mathcal{M}$. To prove the argument on $m$, first consider the linear system $\mathbf{v} = \mathbf{A}\mathbf{u}$ for the $m \times L$ matrix $\mathbf{A}$ of (11). In addition, assume $f_s = f_p = B$. Substituting these values into (10),(12) and using $f_{\text{NYQ}} \geq 2NB$ gives $L > 2N$, namely $\mathbf{A}$ has more than $2N$ columns.

If $m < 2N$, then since $\text{rank}(\mathbf{A}) \leq m$ there exist two $N$-sparse vectors $\bar{\mathbf{u}}_1 \neq \bar{\mathbf{u}}_2$ such that $\mathbf{A}\bar{\mathbf{u}}_1 = \mathbf{A}\bar{\mathbf{u}}_2$. The proof now follows from the following construction. For a given $N$-sparse vector $\mathbf{u}$, choose a frequency interval $\Delta \subset \mathcal{F}_p$ of length $B/2$. Construct a vector $\mathbf{z}(f)$ of spectrum slices, by letting $\mathbf{z}(f) = \mathbf{u}$ for every $f \in \Delta$, and $\mathbf{z}(f) = \mathbf{0}$ otherwise. Clearly, that $\mathbf{z}(f)$ corresponds to some $x(t) \in \mathcal{M}$ (see below an argument that treats the case that this construction results in a complex-valued $x(t)$). Follow this argument for $\bar{\mathbf{u}}_1, \bar{\mathbf{u}}_2$ to provide $\bar{x}_1(t) \neq \bar{x}_2(t)$ within $\mathcal{M}$. Since $\mathbf{A}\bar{\mathbf{u}}_1 = \mathbf{A}\bar{\mathbf{u}}_2$, both $\bar{x}_1(t), \bar{x}_2(t)$ are mapped to the same samples. It can be verified that since $c_{il} = c_{i,-l}^*$, the existence of complex-valued $\bar{x}_1(t) \neq \bar{x}_2(t)$ implies the existence of a corresponding real-valued pair of signals within $\mathcal{M}$, which have the same samples.

The condition (22) comes from the structure of $\mathbf{F}$. For $M < M_{\min}$, $\mathbf{F}$ contains identical columns, for example $\mathbf{F}_1 = \mathbf{F}_{M+1}$. Now, set $\hat{\mathbf{u}}_1$ to be the zero vector except the value $1/d_1$ on the first entry. Similarly, let $\hat{\mathbf{u}}_2$ have zeros except for $1/d_{M+1}$ on the $(M+1)$th entry. We can then use the arguments above to construct the signals $\hat{x}_1(t), \hat{x}_2(t)$ from $\hat{\mathbf{u}}_1, \hat{\mathbf{u}}_2$. It is easy to see that the signals (or their real-valued counterparts) are mapped to the same samples although they are different.

The proof on the necessity of $m \geq 2N$, $M \geq M_{\min}$ for $f_s > f_p$ follows from the same arguments. ∎

We point out that the necessary conditions on $m, M$ may change with other choices of $f_p$. However, $f_p = B$ is sufficient for our purposes, and allows to reduce the total sampling rate as low as possible. In addition, note that it is recommended (though not necessary) to have $M \leq 2^{m-1}$. This requirement stems from the fact that $\mathbf{S}$ is defined over a finite alphabet $\{+1, -1\}$ and thus cannot have more than $2^{m-1}$ linearly independent columns. Therefore, in a sense, the degrees of freedom in $\mathbf{A} = \mathbf{SFD}$ are decreased[2] for $M > 2^{m-1}$. We next show that the conditions of Theorem 1 are also sufficient for blind recovery, under additional conditions.

*Theorem 2 (Sufficient conditions):* Let $x(t)$ be an arbitrary signal within the multiband model $\mathcal{M}$, which is sampled according to Fig. 3 with sign waveforms $p_i(t)$. If

---

[2]Note that repeating the arguments of the proof for $M > 2^{m-1}$ allows to construct spectrum slices $\mathbf{z}(f)$ in the null space of $\mathbf{SF}$. However, these do not necessarily correspond to $x(t) \in \mathcal{M}$ and thus this requirement is only a recommendation.

TABLE I
POSSIBLE PARAMETER CHOICES FOR MULTIBAND SAMPLING

| Model | |
|---|---|
| $N = 6$   $B = 50$ MHz   $f_{\text{NYQ}} = 10$ GHz | |

| Sampling parameters | |
|---|---|
| **Option A** | **Option B** |
| $f_p = \frac{f_{\text{NYQ}}}{195} \approx 51.3$ MHz | $f_p = \frac{f_{\text{NYQ}}}{195} \approx 51.3$ MHz |
| $f_s = f_p \approx 51.3$ MHz | $f_s = 5f_p \approx 256.4$ MHz |
| $m \geq 2N = 12$ | $m \geq \lceil \frac{2N}{5} \rceil = 3$ |
| $M = M_{\min} = 195$ | $M = 195$ |
| $M_{\min} = L = 195$ | $M_{\min} = 195$, $L = 199$ |
| Rate $mf_s \geq 615$ MHz | Rate $mf_s \geq 770$ MHz |

1. $f_s \geq f_p \geq B$, and $f_s/f_p$ is not too large (see the proof);
2. $M \geq M_{\min}$, where $M_{\min}$ is defined in (22);
3. $m \geq N$ for non-blind reconstruction or $m \geq 2N$ for blind;
4. every $2N$ columns of $\mathbf{SF}$ are linearly independent,

then, for every $f \in \mathcal{F}_s$, the vector $\mathbf{z}(f)$ is the unique $N$-sparse solution of (19).

*Proof:* The choice $f_p \geq B$ ensures that every band can contribute only a single non-zero value to $\mathbf{z}(f)$. Fig. 4 and the earlier explanations provide a proof of this statement. As a consequence, $\mathbf{z}(f)$ is $N$-sparse for every $f \in \mathcal{F}_s$.

For $M \geq L$, $\mathbf{D}$ contains nonzero diagonal entries, since $d_l = 0$ only for $l = \pm kM$ for some $k \geq 1$. The same also holds for $M_{\min} \leq M < L$ as long as the ratio $f_s/f_p$ is less than $(M_{\min} + 1)/2$. This implies that $\mathbf{D}$ is nonsingular and rank($\mathbf{SFD}$) = rank($\mathbf{SF}$). Thus linear independence of any column subset of $\mathbf{SF}$ implies corresponding linear independence for $\mathbf{SFD}$.

In the non-blind setting, the band locations imply the support supp($\mathbf{z}(f)$) for every $f \in \mathcal{F}_s$. The other two conditions (on $m$, $\mathbf{SF}$) ensure that (19) can be inverted on the proper column subset, thus providing the uniqueness claim. A closed-form expression is given in (29) below.

In blind recovery, the nonzero locations of $\mathbf{z}(f)$ are unknown. We therefore rely on the following result from the CS literature: A $K$-sparse vector $\mathbf{u}$ is the unique solution of $\mathbf{v} = \mathbf{A}\mathbf{u}$ if every $2K$ columns of $\mathbf{A}$ are linearly independent [28]. This condition translates into $m \geq 2N$ and the condition on $\mathbf{SF}$ of the theorem. ∎

To reduce the sampling rate to minimal we may choose $f_s = f_p = B$ and $m = 2N$ (for the blind scenario). This translates to an average sampling rate of $2NB$, which is the lowest possible for $x(t) \in \mathcal{M}$ [5]. Table I presents two parameter choices for a representative signal model. Option A in the table uses $f_s = f_p$ and leads to a sampling rate as low as 615 MHz, which is slightly above the minimal rate $2NB$ = 600 MHz. Option B is discussed in the next section.

Recall the proof of Theorem 1, which shows that $\mathbf{A}$ has $L > 2N$ columns. Therefore, if $m = 2N$ is sufficiently small, then the requirement $M \geq L$ may contradict the recommendation $M \leq 2^{m-1}$. This situation is rare due to the exponential nature of the upper bound; it does not happen in the examples of Table I. Nonetheless, if it happens, then we may view $x(t) \in$ $\mathcal{M}$ as conceptually having $\rho N$ bands, each of width $B/\rho$, and set $f_p = B/\rho$. The upper bound on $M$ grows exponentially with $\rho$ while the lower bound grows only linearly, thus for some integer $\rho \geq 1$ we may have a valid selection for $M$. This approach requires $m = 2\rho N$ branches which correspond to a large number of sampling channels. Fortunately, this situation can be solved by trading the number of sampling channels for a higher sampling rate $f_s$.

To complete the sampling design, we need to specify how to select the matrix $\mathbf{S}$, namely the sign patterns $\{\alpha_{ik}\}$, such that the last condition of Theorem 2 holds. This issue is shortly addressed in Section IV.

### D. Trading channels for sampling rate

The burden on hardware implementation is highly impacted by the total number of hardware devices, which includes the mixers, the lowpass filters and the ADCs. Clearly, it would be beneficial to reduce the number of channels as low as possible. We now examine a method which reduces the number of channels at the expense of a higher sampling rate $f_s$ in each channel and additional digital processing.

Suppose $f_s = qf_p$, with odd $q = 2q' + 1$. To analyze this choice, consider the $i$th channel of (11) for $f \in \mathcal{F}_p$:

$$\begin{aligned} y_i(f + kf_p) &= \sum_{l=-\infty}^{\infty} c_{il} X(f + kf_p - lf_p) \\ &= \sum_{l=-L_0-k}^{+L_0-k} c_{i,(l+k)} X(f - lf_p) \\ &= \sum_{l=-L_0}^{+L_0} c_{i,(l+k)} X(f - lf_p) \end{aligned} \quad (23)$$

where $-q' \leq k \leq q'$. The first equality follows from a change of variable, and the second from the definition of $L_0$ in (10), which implies that $X(f - lf_p) = 0$ over $f \in \mathcal{F}_p$ for every $|l| > L_0 - q'$. Now, according to (23), a system with $f_s = qf_p$ provides $q$ equations on $\mathcal{F}_p$ for each physical channel. Equivalently, $m$ hardware branches (including all components) amounts to $mq$ channels having $f_s = f_p$. Eq. (24) expands this relation.

Theorem 2 ensures that $\mathbf{z}(f)$ has $N$ nonzero elements for every $f \in \mathcal{F}_s$. Nonetheless, as detailed in the next section, for efficient recovery it is more interesting to determine the joint sparsity level of $\mathbf{z}(f)$ over $\mathcal{F}_s$. As Fig. 4 depicts, over $f \in \mathcal{F}_p$, $\mathbf{z}(f)$ is $2N$-jointly sparse, whereas over the wider range $f \in \mathcal{F}_s$, $\mathbf{z}(f)$ may have a larger joint support set. It is therefore beneficial to truncate the sequences appearing in (23) to the interval $\mathcal{F}_p$, prior to reconstruction. In terms of digital processing, the left-hand-side of (24) is obtained from the input sequence $y_i[n]$ as follows. For every $-q' \leq k \leq q'$, the frequency shift $y_i(f + kf_p)$ is carried out by time modulation. Then, the sequence is lowpass filtered by $h_D[n]$ and decimated by $q$. The filter $h_D[n]$ is an ideal lowpass filter with digital cutoff $\pi/q$, where $\pi$ corresponds to half of the input sampling

$$\begin{pmatrix} y_i(f - q'f_p) \\ \vdots \\ y_i(f) \\ \vdots \\ y_i(f + q'f_p) \end{pmatrix} = \begin{bmatrix} c_{i,L_0-q'} & & \cdots & & & & c_{i,-L_0-q'} \\ \vdots & \ddots & & & & & \vdots \\ c_{i,L_0} & \cdots & c_{i,-1} & c_{i,0} & c_{i,1} & \cdots & c_{i,-L_0} \\ \vdots & & & & & \ddots & \vdots \\ c_{i,L_0+q'} & & \cdots & & & & c_{i,-L_0+q'} \end{bmatrix} \begin{pmatrix} | \\ \mathbf{z}(f) \\ | \end{pmatrix}, \quad f \in \mathcal{F}_p. \quad (24)$$

rate $f_s$. This processing yields the rate $f_p = f_s/q$ sequences

$$\begin{aligned} \tilde{y}_{i,k}[\tilde{n}] &= \left( y_i[n] e^{-j2\pi k f_p n T_s} \right) * h_D[n] \big|_{n=\tilde{n}q} \\ &= \left( y_i[n] e^{-j\frac{2\pi}{q} kn} \right) * h_D[n] \big|_{n=\tilde{n}q}. \end{aligned} \quad (25)$$

Conceptually, the sampling system consists of $mq$ channels which generate the sequences (25) with $f_s = f_p$.

Table I presents a parameter choice, titled Option B, which makes use of this strategy. Thus, instead of the proposed setting of Theorem 2 with $m \geq 12$ channels, uniqueness can be guaranteed from only 3 channels. Observe that the lowest sampling rate in this setting is higher than the minimal $2NB$, since the strategy expands each channel to an integer number $q$ of sequences. In the example, 3 channels are digitally expanded to $3q = 15$ channels. In Section V-C we demonstrate this approach empirically using a finite impulse response (non-ideal) filter to approximate $h_D[n]$.

Theoretically, this strategy allows to collapse a system with $m$ channels to a single channel with sampling rate $f_s = m f_p$. However, each channel requires $q$ digital filters to reduce the rate back to $f_p$, which increases the computational load. In addition, as $q$ grows, approximating a digital filter with cutoff $\pi/q$ requires more taps.

## IV. RECONSTRUCTION

We now discuss the reconstruction stage, which takes the $m$ sample sequences $y_i[n]$ (or the $mq$ decimated sequences $\tilde{y}_{i,k}[\tilde{n}]$) and recovers the Nyquist rate sequence $x(nT)$ (or its analog version $x(t)$). As we explain, the reconstruction also allows to output digital lowrate sequences that captures the information in each band.

Recovery of $x(t)$ from the sequences $y_i[n]$ boils down to recovery of the sparsest $\mathbf{z}(f)$ of (11) for every $f \in \mathcal{F}_s$. The system (11) falls into a broader framework of sparse solutions to a parameterized set of linear systems, which was studied in [11]. In the next subsection we review the relevant results. We then specify them to the multiband scenario.

### A. IMV model

Let $\mathbf{A}$ be an $m \times M$ matrix with $m < M$. Consider a parameterized family of linear systems

$$\mathbf{v}(\lambda) = \mathbf{A}\mathbf{u}(\lambda), \quad \lambda \in \Lambda, \quad (26)$$

indexed by a fixed set $\Lambda$ that may be infinite. Let $\mathbf{u}(\Lambda) = \{\mathbf{u}(\lambda) : \lambda \in \Lambda\}$ be a collection of $M$-dimensional vectors that solves (26). We will assume that the vectors in $\mathbf{u}(\Lambda)$ are jointly $K$-sparse in the sense that $|\text{supp}(\mathbf{u}(\Lambda))| \leq K$. In other words, the nonzero entries of each vector $\mathbf{u}(\lambda)$ lie within a set of at most $K$ indices.

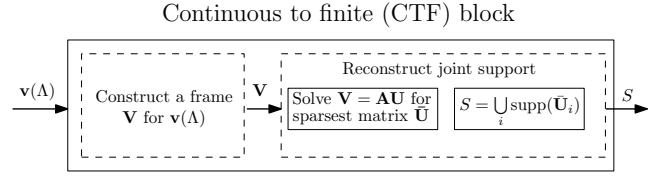

Fig. 5. Recovery of the joint support $S = \text{supp}(\mathbf{u}(\Lambda))$.

When the support $S = \text{supp}(\mathbf{u}(\Lambda))$ is known, recovering $\mathbf{u}(\Lambda)$ from the known vector set $\mathbf{v}(\Lambda) = \{\mathbf{v}(\lambda) : \lambda \in \Lambda\}$ is possible if the submatrix $\mathbf{A}_S$, which contains the columns of $\mathbf{A}$ indexed by $S$, has full column rank. In this case,

$$\mathbf{u}_S(\lambda) = \mathbf{A}_S^\dagger \mathbf{v}(\lambda) \quad (27a)$$
$$\mathbf{u}_i(\lambda) = 0, \quad i \notin S \quad (27b)$$

where $\mathbf{u}_S(\lambda)$ contains only the entries of $\mathbf{u}(\lambda)$ indexed by $S$ and $\mathbf{A}_S^\dagger = (\mathbf{A}_S^H \mathbf{A}_S)^{-1} \mathbf{A}_S^H$ is the (Moore-Penrose) pseudoinverse of $\mathbf{A}_S$. For unknown support $S$, (26) is still invertible if $K = |S|$ is known, and every set of $2K$ columns from $\mathbf{A}$ is linearly independent [11], [28], [29]. In general, finding the support of $\mathbf{u}(\Lambda)$ is NP-hard because it may require a combinatorial search. Nevertheless, recent advances in compressive sampling and sparse approximation delineate situations where polynomial-time recovery algorithms correctly identify $S$ for finite $\Lambda$. This challenge is referred to as a multiple measurement vectors (MMV) problem [27], [29]–[34].

The sparsest solution of a linear system, for unknown support $S$, has no closed-form solution. Thus, when $\Lambda$ has infinite cardinality, referred to as the infinite measurement vectors (IMV) problem [11], solving for $\mathbf{u}(\Lambda)$ conceptually requires an independent treatment for infinitely many systems [11]. To avoid this difficulty of IMV, we proposed in [5], [11] a two step flow which recovers the support set $S$ from a finite-dimensional system, and then uses (27) to recover $\mathbf{u}(\Lambda)$. The algorithm begins with the construction of a (finite) frame $\mathbf{V}$ for $\mathbf{v}(\Lambda)$. Then, it finds the (unique) solution $\bar{\mathbf{U}}$ to the MMV system $\mathbf{V} = \mathbf{A}\mathbf{U}$ that has the fewest nonzero rows. The main result is that $S = \text{supp}(\mathbf{u}(\Lambda))$ equals $\text{supp}(\bar{\mathbf{U}})$, namely the index set of the nonidentically zero rows of $\bar{\mathbf{U}}$. In other words, the support recovery is accomplished by solving only a finite dimensional problem. These operations are grouped in a block entitled continuous to finite (CTF), depicted in Fig. 5. The tricky part of the CTF is in exchanging the infinite IMV system (26) by a finite dimensional one. Computing the frame $\mathbf{V}$, which theoretically involves the entire set $\mathbf{v}(\Lambda)$ of infinitely many vectors, can be implemented straightforwardly in an analog setting as we discuss in the next subsection. Isolating the infiniteness to the frame construction stage enables us to solve (26) exactly with only one finite-dimensional CS problem.

### B. Multiband reconstruction

We now specify the CTF block in the context of multiband reconstruction from the MWC samples. The linear system (11) clearly obeys the IMV model with $\Lambda = \mathcal{F}_s$. In order to use the CTF, we need to construct a frame $\mathbf{V}$ for the measurement set $\mathbf{y}(\Lambda)$. Such a frame can be obtained by computing [11]

$$\mathbf{Q} = \int_{f \in \mathcal{F}_s} \mathbf{y}(f)\mathbf{y}^H(f)\mathrm{d}f = \sum_{n=-\infty}^{+\infty} \mathbf{y}[n]\mathbf{y}^T[n], \qquad (28)$$

where $\mathbf{y}[n] = [y_1[n], \cdots, y_m[n]]^T$ is the vector of samples at time instances $nT_s$. Then, any matrix $\mathbf{V}$, for which $\mathbf{Q} = \mathbf{V}\mathbf{V}^H$, is a frame for $\mathbf{y}(\mathcal{F}_s)$ [11]. The CTF block, Fig. 5, can then be used to recover the support $S = \mathrm{supp}(\mathbf{z}(\mathcal{F}_p))$.

The frame construction (28) is theoretically noncausal. However, $\mathrm{rank}(\mathbf{V}) \leq 2N$ due to the sparsity prior [5], and thus there is no need to collect more than $2N$ linearly independent terms in (28). In practice, only pathological signals would require significantly larger amount of samples to reach the maximal rank [5]. Section V-A demonstrates recovery de-facto from frame construction over a short time interval. Therefore, the infinite sum in (28) can be replaced by a finite sum and still lead to perfect recovery since the signal space is directly identified.

Once $S$ is found,

$$\mathbf{z}_S[n] = \mathbf{A}_S^\dagger \mathbf{y}[n] \qquad (29a)$$
$$z_i[n] = 0, \quad i \notin S, \qquad (29b)$$

where $\mathbf{z}[n] = [z_1[n], \cdots, z_L[n]]^T$ and $z_i[n]$ is the inverse-DTFT of $z_i(f)$. Therefore, the sequences $z_i[n]$ are generated at the input rate $f_s$. At this point, we may recover $x(t)$ by either of the two following options. If $f_{\mathrm{NYQ}}$ is not prohibitively large, then we can generate the Nyquist rate sequences $x(nT)$ digitally and then use an analog lowpass (with cutoff $1/2T$) to recover $x(t)$. The digital sequence $x(nT)$ is generated by shifting each spectrum slice $z_i(f)$ to the proper position in the spectrum, and then summing up the contributions. In terms of digital processing, the sequences $z_i[n]$ are first zero padded:

$$\tilde{z}_i[\tilde{n}] = \begin{cases} z_i[n] & \tilde{n} = nL, n \in \mathbb{Z} \\ 0 & \text{otherwise.} \end{cases} \qquad (30)$$

Then, $\tilde{z}_i[\tilde{n}]$ is interpolated to the Nyquist rate, using an ideal (digital) filter. Finally, the interpolated sequences are modulated in time and summed:

$$x[n] = x(nT) = \sum_{i \in S} (\tilde{z}_i[n] * h_I[n]) e^{2\pi i f_p nT}. \qquad (31)$$

The alternative option is to handle the sequences $z_i[n]$ directly by analog hardware. Every $z_i[n]$ passes through an analog lowpass filter $h_I(t)$ with cutoff $f_s/2$ and gives (the complex-valued) $z_i(t)$. Then,

$$x(t) = \sum_{i \in S, i \geq 0} \mathcal{R}\{z_i(t)\} \cos(2\pi i f_p t) + \mathcal{I}(z_i(t)) \sin(2\pi i f_p t), \qquad (32)$$

where $\mathcal{R}(\cdot), \mathcal{I}(\cdot)$ denote the real and imaginary part of their argument, respectively. By abuse of notation, in both (31) and (32), the sequences $\tilde{z}_i[n]$ are enumerated $-L_0 \leq i \leq L_0$ to shorten the formulas. We emphasize that although the analysis of Section III-B was carried out in the frequency domain, the recovery of $x(t)$ is done completely in the time-domain, via (28)-(32).

The next section summarizes the recovery flow and its advantages from a high-level viewpoint.

### C. Architecture and advantages

Fig. 6 depicts a high-level architecture of the entire recovery process. The sample sequences entering the digital domain are expanded by the factor $q = f_s/f_p$ (if needed). The controller triggers the CTF block on initialization and when identifying that the spectral support has changed. Spectral changes are detected either by a high-level application layer, or by a simple technique discussed hereafter. The digital signal processor (DSP) treats the samples, based on the recovered support, and outputs a lowrate sequence for each active spectrum slice, namely those containing signal energy. A memory unit stores input samples (about $2N$ instances of $\mathbf{y}[n]$), such that in case of a support change, the DSP produces valid outputs in the period required for the CTF to compute the new spectral support. An analog back-end interpolates the sequences and sums them up according to (32). The controller has the ability to selectively activate the digital recovery of any specific band of interest, and in particular to produce an analog counterpart (at baseband) by overriding the relevant carrier frequencies.

**CTF and sampling rate.** The frame construction step of the CTF conceptually merges the infinite collection $\mathbf{z}(\mathcal{F}_s)$ to a finite basis or frame, which preserves the original support. For the CTF to work in the multiband reconstruction, the sampling rate must be doubled due to a specific property that this scenario exhibits. Observe that under the choices of Theorem 2, $\mathbf{z}(\mathcal{F}_p)$ is jointly $2N$-sparse, while each $\mathbf{z}(f)$ is $N$-sparse. This stems from the continuity of the bands which permits each band to have energy in (at most) two spectrum pieces within $\mathcal{F}_p$. Therefore, when aggregating the frequencies the support $\mathrm{supp}(\mathbf{z}(\mathcal{F}_p))$ cannot contain more than $2N$ indices. An algorithm which makes use of several CTF instances and gains back this factor was proposed in [5]. Although the same algorithm applies here as well, we do not pursue this direction so as to avoid additional digital computations.

**MMV recovery complexity.** The CTF block requires solving an MMV system, which is a known NP-hard problem. In practice, sub-optimal polynomial-time CS algorithms may be used for this computation [11], [29], [32]–[34]. The price for tractability is an increase in the sampling rate. In the next section, we quantify this effect for a specific recovery approach. We refer the reader to [29], [33]–[35] for theoretical guarantees regarding MMV recovery algorithms.

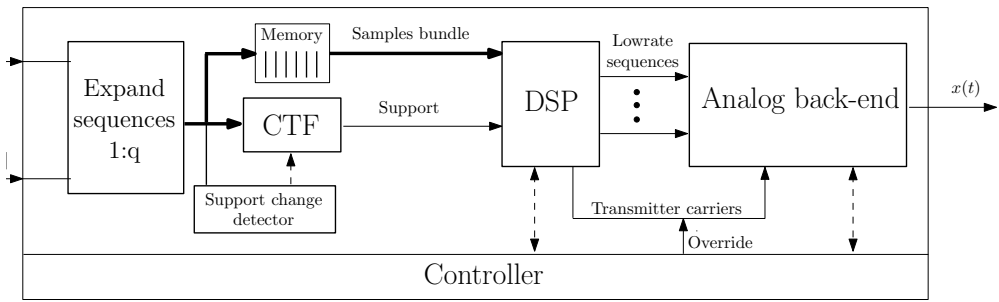

Fig. 6. High-level architecture for efficient multiband reconstruction.

**Realtime processing.** Standard CS algorithms, for the finite $\Lambda$ scenario, couple the tasks of support recovery and the construction of the entire solution. In the infinite scenario, however, the separation between the two tasks has a significant advantage. The support recovery step yields an MMV system, whose dimensions are $m \times L$. Thus, we can control the recovery problem size by setting the number of channels $m$, and setting $L$ via $f_p, f_s$ in (12). Once the support is known, the actual recovery has a closed form (29), and can be carried out in realtime. Indeed, even the recovery of the Nyquist rate sequence (30)-(32), can be done at a constant rate. Had these tasks been coupled, the reconstruction stage would have to recover the Nyquist rate signal directly. In turn, the CS algorithm would have to run on a huge-scale system, dictated by the ambient Nyquist dimension, which is time and memory consuming.

In the context of realtime processing, we comment that the CTF is executed only when the spectral support changes, and thus the short delay introduced by its execution is negligible on average. In a realtime environment, about $2N$ consecutive input vectors $\mathbf{y}[n]$ should be stored in memory, so that in case of a support change the CTF has enough time to provide a new support estimate before the recovery of $\mathbf{z}[n]$, eq. (29), reaches the point that this information is needed. The experiment in Section V-D demonstrates such a realtime solution. In either case, there is no need to recover the Nyquist rate signal before a higher application layer can access the digital information.

In order to notice the support changes once they occur, we can either rely an indication from the application layer, or automatically identify the spectral variation in the sequences $\mathbf{z}[n]$. To implement the latter option, let $S = \text{supp}(\bar{\mathbf{U}})$ be the last support estimate of the CTF, and define $\tilde{S} = S \bigcup \{i\}$ for some entry $i \notin S$. Now, monitor the value of the sequence $z_i[n]$. As long as the support $S$ does not change, the sparsity of $\mathbf{z}[n]$ implies that $z_i[n] = 0$ or contains only small values due to noise. Whenever, this sequence crosses a threshold (for certain number of consecutive time instances) trigger the CTF to obtain a new support estimate. Note that the recovery of $z_i[n]$ requires to implement only one row from $\mathbf{A}_{\tilde{S}}^\dagger$. Since, the values are not important for the detection purpose, the multiplication can be carried out at a low resolution.

**Robustness and sensitivity.** The entire system, sampling and reconstruction, is robust against inaccuracies in the parameters $f_s, f_p$. This is a consequence of setting the parameters according to Theorem 2, with only the inequalities $f_s \geq f_p \geq B$. In particular, $f_p$ is chosen above the minimal to ensure safety guard regions against hardware inaccuracies or signal mismodeling. Furthermore, observe that the exact values of $f_s, f_p$ do not appear anywhere in the recovery flow: the expanding equations (25), the frame construction (28), the CTF block – Fig. 5, and the recovery equations (29). Only the ratio $q = f_s/f_p$ is used, which remains unchanged if the a single clock circuitry is used in the design. In addition, in the recovery of the Nyquist rate sequence (31), only the ratio $T/T_p$ is used, which remains fixed for the same reasons. When recovering $x(t)$ via (32), $f_p$ is provided to the back-end from the same clock triggering the sampling stage. The recovery is also stable in the presence of noise as numerically demonstrated in Section V-A.

**Digital implementation.** The sample vectors $\mathbf{y}[n]$ arrive synchronously to the digital domain. As mentioned earlier, a possible interface is to trigger a digital processer from the same clock driving the ADCs, namely at rate $1/T_s$. Since the digital input rate is relatively low, on the order of $B$ Hz, commercial cheap DSPs can be used. However, here the actual number of channels $m$ has a great impact. Each sample is quantized by the ADC to a certain number of bits, say 8 or 16. The bus width towards the DSP becomes of length $8m$ or $16m$, respectively. Care must be taken when choosing the processing unit in order to accommodate the bus width. Note that some recent DSPs have analog inputs with built-in synchronized ADCs so as to avoid such a problem. See other aspects of quantization in Section V-E.

Finally we point out an advantage with respect to the reconstruction of a multicoset based receiver. The IMV formulation holds for this strategy with a different sampling matrix $\mathbf{A}$ [5]. However, the IMV system requires a (Nyquist rate) zero padded version of (2) in this case. Consequently, constructing a frame $\mathbf{V}$ from the multicoset low-rate sequences (2) requires interpolating the sample sequences to the Nyquist rate. Only then can $\mathbf{Q}$ be computed (see (61)-(62) in [5]). Furthermore, reconstruction of the signal $x(t)$ also requires the same interpolation to the Nyquist grid, that is even for a known spectral support. In contrast, the current mixing stage has the advantage that the IMV is expressed directly in terms of the lowrate sequences $y_i[n]$, and the computation of $\mathbf{Q}$ in (28) is carried out directly on the input sequences. In fact, one may implement an adaptive frame construction at the input rate $f_s$. Digital processing at rate $f_s$ is obviously preferred over a processor running at the Nyquist rate.

### D. Choosing the sign patterns

Theorem 2 requires that for uniqueness, every $2N$ columns of $\mathbf{SF}$ must be linearly independent. To apply the CTF block the requirement is strengthened to every $4N$ columns, which also implies the minimal number of rows in $\mathbf{S}$ [5]. Verifying that a set of sign patterns $\{\alpha_{ik}\}$ satisfies such a condition is computationally difficult because one must check the rank of every set of $4N$ columns from $\mathbf{SF}$. In practice, when noise is present or when solving the MMV by sub-optimal polynomial-time CS algorithms, the number of rows in $\mathbf{S}$ should be increased beyond $m = 4N$. A preliminary discussion on the required dimensions of $\mathbf{S}$ is quoted below from the conference version of this work [36]. The actual choice of the patterns will be investigated in future work.

Consider the system $\mathbf{v} = \mathbf{Au}$, where $\mathbf{u}$ is an unknown sparse vector, $\mathbf{v}$ is the measurement vector, and $\mathbf{A}$ is of size $m \times M$. A matrix $\mathbf{A}$ is said to have the restricted isometry property (RIP) [27] of order $K$, if there exists $0 \le \delta_K < 1$ such that

$$(1-\delta_K)\|\mathbf{u}\|^2 \le \|\mathbf{Au}\|^2 \le (1+\delta_K)\|\mathbf{u}\|^2 \qquad (33)$$

for every $K$-sparse vector $\mathbf{u}$ [27]. The requirement of Theorem 2 thus translates to $\delta_{2N} < 1$. The RIP requirement is also hard to verify for a given matrix. Instead, it can be easier to prove that a random $\mathbf{A}$, chosen from some distribution, has the RIP with high probability. In particular, it is known that a random sign matrix, whose entries are drawn independently with equal probability, has the RIP of order $K$ if $m \ge CK\log(M/K)$, where $C$ is a positive constant independent of everything [37]. The log factor is necessary [38]. The RIP of matrices with random signs remains unchanged under any fixed unitary transform of the rows [37]. This implies that if $\mathbf{S}$ is a random sign matrix, possibly implemented by a length $M$ shift register per channel, then $\mathbf{SF}$ has the RIP of order $2N$ for the above dimension selection. Note that $\mathbf{D}$ is ignored in this analysis, since the diagonal has nonzero entries and thus $\mathrm{supp}(\mathbf{Du}) = \mathrm{supp}(\mathbf{u})$ for any vector $\mathbf{u}$. In particular, it is known that recovery using the program (35) below depends only on the signs of the nonzero values of $\mathbf{u}$, which are unchanged under diagonal scaling.

To proceed, observe that solving for $\mathbf{u}$ would require the combinatorial search implied by

$$\min_{\mathbf{u}} \|\mathbf{u}\|_0 \text{ s.t. } \mathbf{v} = \mathbf{Au}. \qquad (34)$$

A popular approach is to approximate the sparsest solution by

$$\min_{\mathbf{u}} \|\mathbf{u}\|_1 \text{ s.t. } \mathbf{v} = \mathbf{Au}. \qquad (35)$$

The relaxed program, named basis pursuit (BP) [39], is convex and can be tackled with polynomial-time solvers [27]. Many works have analyzed the basis pursuit method and its ability to recover the sparsest vector $\mathbf{u}$. For example, if $\delta_{2K} \le \sqrt{2}-1$ then (35) recovers the sparsest $\mathbf{u}$ [40]. The squared error of the recovery in the presence of noise or model mismatch was also shown to be bounded under the same condition [40]. Similar conditions were shown to hold for other recovery algorithms. In particular, [35] proved a similar argument for a mixed $\ell_2/\ell_1$ program in the MMV setting (which incorporates the joint sparsity prior). See also [34].

In practice, the matrix $\mathbf{S}$ is not random once the sampling stage is implemented, and its RIP constant cannot be calculated efficiently. A reviewer also pointed out that when implementing a binary sequence using feedback logic, as popular for m-sequences, the set of possible sign patterns is much smaller than $2^M$. In this setting, alternative randomness properties, such as almost $k$-wise independency can be beneficial [41]. Extensive simulations on synthesized data are often used to evaluate the performance and the stability of a CS system when RIP values are difficult to compute (*e.g.,* see [11], [29], [31]). Clearly, the numerical results do not ensure a desired RIP constant. Nonetheless, for practical applications, the behavior observed in simulations may be sufficient. The discussion above implies that stable recovery of the MMV of Fig. 5 requires roughly

$$m \approx 4N\log(M/2N) \qquad (36)$$

channels to estimate the correct support, using polynomial-time algorithms.

### V. NUMERICAL SIMULATIONS

We now demonstrate several engineering aspects of our system, using numerical experiments:

1. A wideband design example in the presence of wideband noise, for a synthesized signal with rectangular transmission shapes;
2. Hardware simplifications: using a single shift-register to implement several periodic waveforms $p_i(t)$ at once;
3. Collapsing the number of hardware channels, evaluating the idea presented in Section III-D;
4. Fast adaption to time-varying support, for quadrature phase shift keying (QPSK) transmissions;
5. Quantization effects.

### A. Design example

To evaluate the performance of the proposed system (see Fig. 3) we simulate the system on test signals contaminated by white Gaussian noise.

More precisely, we evaluate the performance on 500 noisy test signals of the form $x(t) + w(t)$, where $x$ is a multiband signal and $w$ is a white Gaussian noise process. The multiband model of Table I is used hereafter. The signal consists of 3 pairs of bands (total $N = 6$), each of width $B = 50$ MHz, constructed using the formula

$$x(t) = \sum_{i=1}^{3} \sqrt{E_i B}\,\mathrm{sinc}(B(t-\tau_i))\cos(2\pi f_i(t-\tau_i)), \qquad (37)$$

where $\mathrm{sinc}(x) = \sin(\pi x)/(\pi x)$. The energy coefficients are $E_i = \{1, 2, 3\}$ and the time offsets are $\tau_i = \{0.4, 0.7, 0.2\}$ $\mu$secs. The exact values $X(f)$ takes on the support do not affect the results and thus $E_i, \tau_i$ are fixed in all our simulations. For every signal the carriers $f_i$ are chosen uniformly at random in $[-f_{\mathrm{NYQ}}/2, f_{\mathrm{NYQ}}/2]$ with $f_{\mathrm{NYQ}} = 10$ GHz.

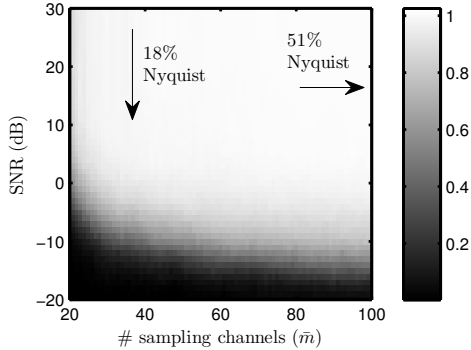

Fig. 7. Image intensity represents percentage of correct support set recovery $\hat{S} = S$, for reconstruction from different number of sampling sequences $\bar{m}$ and under several SNR levels.

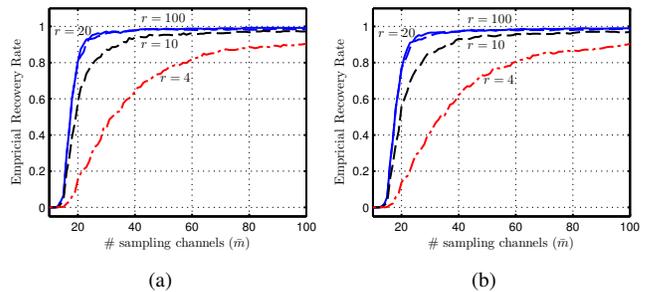

Fig. 8. Percentage of correct support recovery, when drawing the sign patterns randomly only for the first $r$ channels. Results are presented for (a) SNR=25 dB and (b) SNR=10 dB.

We design the sampling stage according to "Option A" of Table I. Specifically, $f_s = f_p = f_{\text{NYQ}}/195 \simeq 51.3$ MHz. The number of channels is set to $m = 100$, where each mixing function $p_i(t)$ alternates sign at most $M = M_{\min} = 195$ times. Each sign $\alpha_{ik}$ is chosen uniformly at random and fixed for the duration of the experiment. To represent continuous signals in simulation, we place a dense grid of 50001 equispaced points in the time interval $[0, 1\mu\text{secs}]$. The time resolution under this choice, $T/5$, is used for accurate representation of the signal after mixing, which is not bandlimited. The Gaussian noise is added and scaled so that the test signal has the desired signal-to-noise ratio (SNR), where the SNR is defined to be $10 \log(\|x\|^2/\|w\|^2)$, with the standard $l_2$ norms. To imitate the analog filtering and sampling, we use a lengthy digital FIR filter followed by decimation at the appropriate factor. After removing the delay caused by this filter, we end up with 40 samples per channel at rate $f_s$, which corresponds to observing the signal for 780 nsecs. We emphasize that these steps are required only when simulating an analog hardware numerically. In practice, the continuous signals pass through an analog filter (*e.g.,* an elliptic filter), and there is no need for decimation or a dense time grid.

The support of the input signal is reconstructed from $\bar{m} \leq m$ channels. (More precisely, $S = \text{supp}(\mathbf{z}(\mathcal{F}_p))$ is recovered.) We follow the procedure described in Fig. 5 to reduce the IMV system (19) to an MMV system. Due to Theorem 2, $\mathbf{Q}$ is expected to have (at most) $2N = 12$ dominant eigenvectors. The noise space, which is associated with the remaining negligible eigenvalues is discarded by simple thresholding ($10^{-9}$ is used in the simulations). Then, the frame $\mathbf{V}$ is constructed and the MMV is solved using simultaneous orthogonal matching pursuit [31], [32]. We slightly modified the algorithm to select a symmetric pair of support indices in every iteration, based on the conjugate symmetry of $X(f)$. Success recovery is declared when the estimated support set is equal the true support, $\hat{S} = S$. Correct recovery is also considered when $\hat{S} \supset S$ contains a few additional entries, as long as the corresponding columns $\mathbf{A}_S$ are linearly independent. As explained, recovery of the Nyquist rate signal can be carried out by (31)-(32). Fig. 7 reports the percentage of correct support recoveries for various numbers $\bar{m}$ of channels and several SNRs.

The results show that in the high SNR regime, correct recovery is accomplished when using $\bar{m} \geq 35$ channels, which amounts to less than 18% of the Nyquist rate. This rate conforms with (36) which predicts an order of $4N \log(M/2N) \simeq 30$ channels for stable recovery. A saving factor 2 is possible if using more than a single CTF block and a complicated processing (see [5] for details) or by brute-force MMV solvers with exponential recovery time. An obvious trend which appears in the results is that the recovery rate is inversely proportional to the SNR level and to the number of channels $\bar{m}$ used for reconstruction.

### B. Simplifying the mixing stage

Each channel needs a mixing function $p_i(t)$, which supposedly requires a shift register of $M$ flip flops. In the setting of Fig. 7, every channel requires $M = 195$ flip flops with a clock operating at $(T_p/M)^{-1} = 10$ GHz.

We propose a simple method to reduce the total number of flip flops by sharing the same register by a few channels, and using consecutive taps to produce several mixing functions simultaneously. This strategy however reduces the degrees of freedom in $\mathbf{S}$ and may affect the recovery performance. To qualitatively evaluate this approach, we generated sign matrices $\mathbf{S}$ whose first $r$ rows are drawn randomly as before. Then, the $i$th row, $r < i \leq m$, is five cyclic shifts (to the right) of the $(i-r)$th row. Fig. 8 reports the recovery success for several choices of $r$ and two SNR levels. As evident, this strategy enables a saving of 80% of the total number of flip flips, with no empirical degradation in performance.

### C. Collapsing analog channels

Section III-D introduced a method to collapse $q$ sampling channels to a single channel with a higher sampling rate $f_s = qf_p$. To evaluate this strategy, we choose the parameter set "Option B" of Table I. Specifically, the system design of Section V-A is now changed to $f_s = 5f_p$, with $m = 20$ physical channels.

In the simulation, the time interval in which the signal is observed is extended to $[0, 4\mu\text{secs}]$, such that every channel records (after filtering and sampling) about 500 samples. The extended window enables accurate digital filtering in order to separate each sequence to $q = 5$ different equations. We design a 100-tap digital FIR filter with the MATLAB command h=fir1(100,1/q) to approximate the optimal filter $h_D[n]$ of Section III-D. Then, for the $i$th sample sequence $y_i[n]$, $h$ is convolved with each of the modulated versions $y_i[n]e^{j2\pi/q\,ln}$, where $-q' \leq l \leq q' = 2$. Fig. 9 reports the recovery performance for different SNR levels and versus the number of sampling channels. The performance trend remains as in Fig. 7. In particular, $35/q = 7$ channels achieve an acceptable recovery rate. This implies a significant saving in hardware components. The combination of collapsing channels and sharing the same shift register for different channels were realized in [12] for $m = 4, M = 96$.

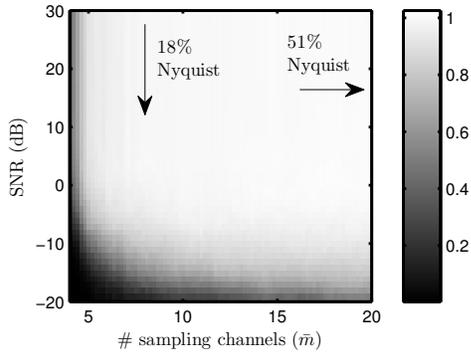

Fig. 9. Image intensity represents percentage of correct support set recovery $\hat{S} = S$, for reconstruction from different number of hardware channels $\bar{m}$ and under several SNR levels. The input sample sequences are expanded to $\bar{m}f_s/f_p$ digital sequences.

### D. Time-varying support

To demonstrate the realtime capabilities of our system, we consider a communication system with 3 concurrent quadrature phase shift keying (QPSK) transmissions of width $B = 30$ MHz each. Each QPSK signal is given by

$$x(t) = \sqrt{\frac{2E_{\text{sym}}}{T_{\text{sym}}}} \left( \sum_n I[n]s(t - nT_{\text{sym}}) \right) \cos(2\pi f_c t) \quad (38)$$
$$+ \left( \sum_n Q[n]s(t - nT_{\text{sym}}) \right) \sin(2\pi f_c t),$$

where $E_{\text{sym}}, T_{\text{sym}} = 2/B$ are the energy and the duration of a symbol. The in-phase and quadrature bit streams are $I[n], Q[n]$, and $s(t)$ is the pulse shaping. We chose the standard shaping $s(t) = \text{sinc}(t/T_{\text{sym}})$ and generated the bit streams uniformly at random. The power spectral density around the carrier $f_c$ is illustrated in Fig. 10-a. Evidently, reallife transmissions have nonsharp edges, as opposed to nice rectangular sinc signals, which were synthesized in (37).

The experiment was set up as follows. Three QPSK signals of the form (38) were generated $x_1(t), x_2(t), x_3(t)$ with symbol energies $1, 2, 3$ respectively. The carriers $f_i$ were drawn as before uniformly at random over a wideband range with $f_{\text{NYQ}} = 10$ GHz. Every $10\mu$secs the carrier were re-drawn independently of their previous values. Each interval of $10\mu$sec gave about 500 time samples $\mathbf{y}[n]$. In addition, the SNR was fixed to 30 dB. The sampling parameters are the same of Fig. 7, except for a fixed number of channels $m = 40$ so as to simplify the presentation.

In order to handle the time-varying support, we decided to use $N_{\text{CTF}} = 50$ time samples $\mathbf{y}[n]$ for the frame construction of the CTF. In addition, we considered the architecture of Fig. 6 with a memory stack that can save only $N_{\text{MEM}} = 20$ vectors. As a result, whenever the spectral support changes, the lowrate sequences $\mathbf{z}[n]$ remain valid only for 20 cycles, and then becomes invalid for 30 more cycles, until the CTF provides a new support estimate. To identify the support changes, we used the technique described earlier in Section IV-C.

Fig. 10-b shows the normalized squared baseband error, which is defined as

$$\text{Baseband error}[n] = \frac{\|\hat{\mathbf{z}}[n] - \mathbf{z}[n]\|^2}{\|\mathbf{z}[n]\|^2}, \quad (39)$$

where $\mathbf{z}[n]$ corresponds to the signal $x(t)$ without noise, according to (13), while $\hat{\mathbf{z}}[n]$ are the actual recovered sequences, including noise and possible wrong indices in the recovered support. We measure the baseband error, rather than the output error $\|\hat{x}(t) - x(t)\|^2/\|x(t)\|^2$, since the lowpass filter in the output recovery, either $h_I[n]$ in (31) or $h_I(t)$ in (32), has its own memory which smooths out the error to negligible values. In the figure, the noise floor is due to the normalization in (39) and our choice of 30 dB SNR.

This experiment highlights that the CTF requires only a short duration to estimate the support. Once the new estimate is ready, the baseband error, and consequently reconstruction, are correct. In the experiment, we intentionally used a memory size $N_{\text{MEM}}$ smaller than $N_{\text{CTF}}$, in order to demonstrate error in this setting. In practice, one should use $N_{\text{MEM}} \geq N_{\text{CTF}}$ for normal operation. When changing the SNR and the number of channels, we found that $N_{\text{CTF}}$ can be much lower than 50. The bottom line is that the CTF introduces only a short delay in realtime environments, and the memory requirements are consequently very low.

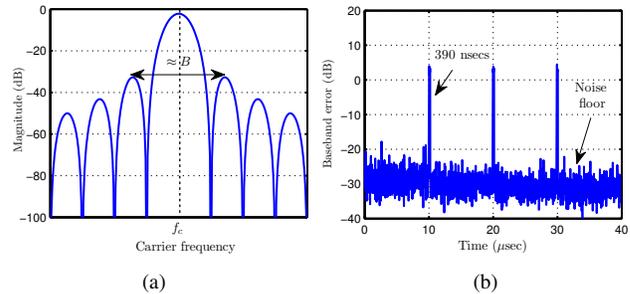

Fig. 10. The spectral density of a QPSK transmission is plotted in (a). Reconstruction of a signal with time-varying spectral support is demonstrated in (b).

### E. Quantization

The ADC device performs two tasks: taking pointwise samples of the input (up to the bandwidth limitation), and quantizing the samples to a predefined number of bits. So far, we have ignored quantization issues. A full study of these effects is beyond the current scope. Nonetheless, we provide

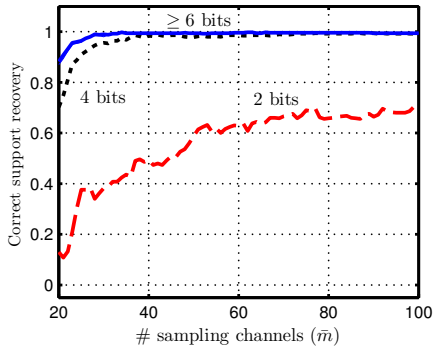

Fig. 11. Support recovery from quantized samples of QPSK transmissions.

a preliminary demonstration of the system capabilities in that context.

Quantization is usually regarded as additive noise at the input, though the noise distribution is essentially different from the standard model of white Gaussian noise. Since Fig. 7 shows robustness to noise, it is expected that the system can handle quantization effects in the same manner. To perform the experiment, we used the setting of the first experiment, Fig. 7, with the following exceptions: QPSK transmissions (37), no additive wideband noise (in order to isolate the quantization effect), and a variable number of bits to represent $\mathbf{y}[n]$. We used the simplest method for quantization – uniformly spaced quantization steps that covers the entire dynamic range of $\mathbf{y}[n]$. Fig. 11 shows that indeed the support recovery functions properly even from a few number of bits.

## VI. DISCUSSION

### A. Related work

The random demodulator is a recent system which also aims at reducing the sampling rate below the Nyquist barrier [9], [10]. The system is presented in Fig. 12. The input signal $f(t)$ is first mixed by a sign waveform with a long period, produced by a pseudorandom sign generator which alternates at rate $W$. The mixed output is then integrated and dumped at a constant rate $R$, resulting in the sequence $y[n]$.

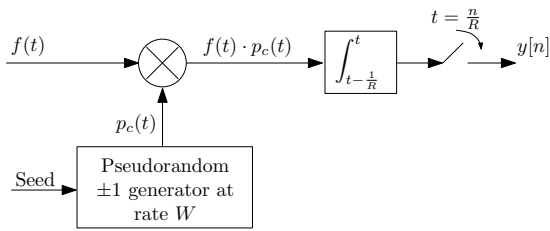

Fig. 12. Block diagram of the random demodulator [10].

The signal model for which the random demodulator was designed consists of multitone functions:

$$f(t) = \sum_{\omega \in \Omega} a_\omega e^{-j2\pi\omega t}, \quad t \in [0,1), \quad (40)$$

where $\Omega$ is a finite set of tones

$$\Omega \subset \{0, \pm 1, \pm 2, \cdots, \pm(W/2-1), W/2\}. \quad (41)$$

The analysis in [10] shows that $f(t)$ can be recovered from $y[n]$, using the linear system

$$\mathbf{y} = \mathbf{\Phi}\mathbf{a}, \quad (42)$$

where $\mathbf{\Phi}$ is $R \times W$ matrix and $\mathbf{a}$ collects the coefficients $a_\omega$.

Despite the somewhat visual similarity between Fig. 12 and Fig. 3, the systems are essentially different in many aspects. The most noticeable is the discrete multitone setting in contrast to the analog multiband model that was considered throughout this paper. When attempting to approximate analog signals in the discrete model, such as those used in the previous section, the number of tones $W$ is about the Nyquist rate, and $R = \text{const} \cdot NB$ is required [10]. In practice, this results in a huge-scale $\mathbf{\Phi}$ (millions of rows by 10's of millions of columns), which may not allow to solve for the coefficients $a_\omega$ in a reasonable amount of computations. In contrast, the MWC is developed for continuous signals, and the matrix $\mathbf{A}$ has low dimensions, $35 \times 195$ in our experiments, for the same signal parameters.

Besides model and computational aspects, the systems also differ in terms of hardware. Our approach is easily adapted to arbitrary periodic waveforms by just re-calculating the Fourier coefficients $c_{il}$ in (7). In contrast, the analysis in [10] is more tailored for the specific choice of sign waveforms. The hardware of [10] also requires accurate integration, as opposed to flexible analog filter design in the MWC.

Finally, we point out that (42) aims at Nyquist rate recovery. In contrast, our approach combines standard sampling theory tools, such as frequency-domain analysis, Section III-B, and incorporates CS only where beneficial. The CS problem of the CTF, eq. (19), is used only for support recovery, which is the key for reducing recovery complexity and allowing low-rate processing. A detailed comparison of our system with the random demodulator appears in [42].

### B. Concluding remarks

We presented a sub-Nyquist sampling system, the modulated wideband converter, which is designed independently of the spectral support of the input signal. The analog front-end supports wideband applications and can also be used to sample wideband inputs occupying the entire spectral support. A unified digital architecture for spectrum-blind reconstruction and for low-rate processing was also provided. The architecture consists of digital support recovery and an analog back-end. The digital operations required for the support recovery need only a small number of observations, thus introducing a short delay. Once the support is known, various realtime computations are possible. Recovery of the original signal at the Nyquist rate is only one application. Perhaps more important is the potential to digitally process any information band at a low rate.

This work bridges theory to practice. In theory, we prove that analog signals are determined from minimal rate samples. In the bridge to practice, we utilized numerical simulations to prove the concept of stable recovery in challenging wideband conditions. Finally, we presented various practical considerations, both for the implementation of the analog front-end (e.g. setting the number of channels, trading system branches by a higher sampling rate, and some potential hardware simplifications), and for the digital stage (e.g. low-rate and realtime processing, handling time-varying spectrum, and quantization). The engineering aspects are the prime focus of the current paper, while future work will sharpen the theoretical understandings and report on circuit-level implementation [12].

The current work embeds theorems and algorithms from compressed sensing (CS), an emerging research field which exploits sparsity for dimension reduction. The mainstream line of CS papers studies sparsity for discrete and finite vectors. The random demodulator expands this approach by parameterizing continuous signals in a finite setting. In contrast, this work continues the line of [5], [11] and belongs to a recently-developed framework within CS [35], [43], [44], which studies signals from a truly continuous domain. Within this analog framework, we propose selecting a practical implementation among the various possible sampling stages covered by [43].

ACKNOWLEDGMENT

The authors would like to thank Prof. Joel A. Tropp for fruitful discussions and for helpful comments on the first draft of this manuscript, and Yilun Chen for insightful discussions regarding the simulations. The authors also appreciate the constructive comments of the the anonymous reviewers.